\documentclass[12pt]{iopart}
\usepackage{iopams}

\newcommand{\beq}{\begin{equation}}
\newcommand{\eeq}{\end{equation}}

\newcommand{\cross}{\mbox{\boldmath $\times$}}
\newcommand{\cendot}{\mbox{\boldmath $\cdot\,$}}
\newcommand{\im}{{\rm i}}

\newcommand{\bfp}{\mbox{\boldmath $p$}}
\newcommand{\bfb}{\mbox{\boldmath $b$}}
\newcommand{\bfn}{\mbox{\boldmath $n$}}
\newcommand{\bfk}{\mbox{\boldmath $k$}}
\newcommand{\bfs}{\mbox{\boldmath $s$}}
\newcommand{\bfl}{\mbox{\boldmath $l$}}
\newcommand{\bfc}{\mbox{\boldmath $c$}}
\newcommand{\bfg}{\mbox{\boldmath $g$}}
\newcommand{\bfE}{\mbox{\boldmath $E$}}
\newcommand{\bfe}{\mbox{\boldmath $e$}}
\newcommand{\bfU}{\mbox{\boldmath $U$}}
\newcommand{\msM}{\mbox{$\mathsf{M}$}}
\newcommand{\msR}{\mbox{$\mathsf{R}$}}
\newcommand{\mcT}{\mbox{$\mathcal{T}$}}
\newcommand{\mcL}{\mbox{$\mathcal{L}$}}
\newcommand{\mcF}{\mbox{$\mathcal{F}$}}
\newcommand{\mcH}{\mbox{$\mathcal{H}$}}
\newcommand{\mcN}{\mbox{$\mathcal{N}$}}

\begin{document}

\title[Global description of light beams and their  geometric phase]{Light beams with general direction and polarization: global description  and  geometric phase}

\author{R~Nityananda$^1$ and S~Sridhar$^2$}

\address{$^1$ TIFR Centre for Interdisciplinary Sciences, 21, Brundavan colony, Narsinghi, Hyderabad~500~089, India\,; National Centre for Radio Astrophysics, TIFR, Pune~411~007, India}

\address{$^2$ Raman Research Institute, Sadashivanagar, Bangalore~560~080, India}

\ead{\mailto{rajaram@ncra.tifr.res.in}, \mailto{ssridhar@rri.res.in}}

\begin{abstract}
We construct the manifold describing the family of plane monochromatic light waves with all directions, polarizations, phases and intensities. A smooth description of polarization, valid over the entire sphere $S^2$ of directions, is given through the construction of an orthogonal basis pair of \emph{complex} polarization vectors for each direction; any light beam is then uniquely and smoothly specified by giving its direction and two complex amplitudes. This implies that the space of all light beams is the six dimensional manifold $\,S^2\times\mathbb{C}^2\,$, the Cartesian product of a sphere and a two dimensional complex vector space. A Hopf map (i.e mapping the two complex amplitudes to the Stokes parameters) then leads to   the four dimensional manifold $\,S^2\times S^2\,$ which describes beams with all directions and polarization states. This product of two spheres can be viewed as an \emph{ordered} pair of two points on a single sphere, in contrast to earlier work in which the same system was represented using Majorana's mapping of the states of a spin one quantum system to an \emph{unordered} pair of points on a sphere. This is a different manifold, $CP^2\,$, two dimensional complex projective space, which does not faithfully represent the full space of all directions and polarizations. Following the now--standard framework, we exhibit the fibre bundle whose total space is the set of all light beams of non--zero intensity, and base space $\,S^2\times S^2\,$. We give the $U(1)$ connection which determines the geometric phase as the line integral of a one--form along a closed curve in the total space. Bases are classified as globally smooth, global but singular, and local, with the last type of basis being defined only when the curve traversed by the system is given. Existing as well as new formulae for the geometric phase are presented in this overall framework. 
\end{abstract}

\submitto{\JPA}
\maketitle

\section{Introduction and summary}
Bertolotti \cite{ber26} formulated the evolution of linear polarization as light traverses a space curve  in  an inhomogeneous but locally isotropic medium in the geometrical optics limit. He concluded that the electric vector is 
parallel--transported with respect to a connection derived from a conformally flat metric, where Euclidean  distances are scaled by the local value of the refractive index. Rytov \cite{ryt38} derived this evolution law independently from a WKB treatment of Maxwell's equations, and also expressed it as a phase difference per unit length between the two circular polarizations, proportional   to the torsion of the space curve. Vladimirskii \cite{vla41}  brought out the following geometrical implication: polarization vectors live in the tangent plane to the sphere of directions and undergo parallel displacement as the direction changes. This implies that after the tangent vector to the curve returns to its original value (e.g. after one turn of a helix), the polarization rotates by an angle equal to the solid angle enclosed by the closed trajectory of the tangent vector on this sphere. 

Pancharatnam \cite{pan56}, in the context of novel interference patterns shown by absorbing biaxial crystals, formulated the phase which now bears his name, equal to one--half of the solid angle traversed on the Poincare sphere which represents polarization states.  The work of Berry \cite{ber84} on the phase change of a  quantum state,   evolving adiabatically as   the Hamiltonian describes a closed path in a parameter space, and its later generalizations, provides  the natural  framework  in which to discuss this class of optical  situations; see e.g. \cite{mv89, sw89} for reviews of early work. The Berry or geometric phase depends on the path traversed in the parameter space (i.e the sphere of directions or of polarizations) but not on the rate of traversal. 

Bhandari \cite{bha89, bha97}, Hannay \cite{han98} and Tavrov et al \cite{tav00}, among others, treated geometric phases under  the simultaneous evolution of direction and polarization. This can be viewed as occurring in a four dimensional space, whose global structure is naturally of interest. Bhandari and Hannay worked with the four dimensional ray space (the space of physical states) of a spin one quantum system. Such a ray space is constructed from a three dimensional complex vector space (Hilbert space) by identifying vectors differing only in normalization and phase; it is denoted by $\,CP^2\,$,  two dimensional complex projective space. The spin one  description is indeed a faithful mapping of the elliptical orbits traversed by the electric fields of the light beams being considered, which are the same as the orbits of a three dimensional isotropic harmonic oscillator. However, two problems prevent 
$\,CP^2\,$ from being the four dimensional space that  faithfully represents all directions and polarizations. One is that a given ellipse traversed by the electric vector in a plane in (real) three dimensional space can correspond to two opposite directions of propagation. This could be resolved by doubling 
$\,CP^2\,$, using one copy for each sense of circular or elliptic polarization.
However, another problem arises on the boundary between these two regions: an electric field  which is linearly polarized  along a given direction can belong, not just to two but, to an entire circle of directions of propagation in a plane perpendicular to it; i.e. the subspace of linear polarizations in the spin one  model is two dimensional, whereas it should really be three dimensional. Thus the correct global nature of the space is worth investigating.  

In this paper we  first determine the  six dimensional manifold which faithfully represents light beams with all directions, polarizations, phases and intensities. As a by--product we  give a direct demonstration that the four dimensional space that represents all directions and polarizations is 
$\,S^2\times S^2\,$, the Cartesian product of two spheres, which is distinct from $\,CP^2\,$. As discussed below this implies more than the known fact that each of direction and polarization forms a sphere, when considered separately. In principle, the total space of interest could have been a non--trivial, 
twisted  $S^2$ (polarization) fibre bundle over $S^2$ (directions) as 
base.\footnote{Incidentally, $CP^2$ cannot be represented as such a bundle.} Given a direction of propagation with unit vector $\bfk$, two \emph{real} orthogonal unit vectors can be chosen as  a pair of linearly polarized  basis states in the tangent plane. Then the complex electric vector, of any general polarization, can be specified by affixing complex amplitudes to the basis vectors. These two complex amplitudes contain information about polarization, phase and intensity. Hence the six dimensional space of all light beams can be thought of as a fibre bundle with base $S^2$ the sphere of directions, and fibre $\mathbb{C}^2$ a two dimensional complex vector space. A smooth extension of the locally constructed real, linearly polarized, basis vectors to all directions $\bfk \in S^2\,$ runs into the obstruction that it is impossible to construct a smooth non--vanishing vector field on $S^2$. We show below that this obstruction can be overcome by choosing a \emph{complex} basis. 

In  section~2 we exhibit such an orthogonal pair of complex basis vectors for each direction which is globally smooth on $S^2$. Section~3 brings out some implications of this construction. The existence of such a globally smooth basis implies that the fibre bundle is trivial. Therefore the six dimensional manifold is $\,S^2\times \mathbb{C}^2\,$. It then follows from the standard transition from complex amplitudes to the Stokes parameters ( the  Hopf map) that the four dimensional manifold which describes beams with all directions and polarizations is $\,S^2\times S^2\;$. Given a general state of polarization with a general direction, one can pick a unique point $\bfk$ on the sphere of directions and, using this basis, assign a unique point $\bfs$ on the Poincar\'e sphere. 

The geometrical phase, discovered by Berry in the context of  
adiabatic evolution of a quantum--mechanical system, was  restated in the language of modern differential geometry by Simon \cite{sim83}, as the 
(an)holonomy of a $U(1)$ connection on a fibre bundle. Samuel and 
Bhandari \cite{sb88} offered a general setting for the geometric phase, 
allowing for comparison of the phase between any two nonorthogonal states in Hilbert space. In section~4 we follow the formulation of \cite{sb88} and construct a fibre bundle, whose total space is the set of all light beams of non--zero intensity $\,S^2\times \mathbb{C}^2\setminus\{{\bf 0}\}\,$, with base space $\,S^2\times S^2\,$. We use Pancharatnam's method of comparing phases of different polarizations and Vladimirskii's notion of parallel displacement   to derive a $U(1)$ connection on the fibre bundle, and give a basis--independent definition of the geometric phase. In section~5 we use the global basis discussed above, and write the geometric phase as the sum of two phases; one which resembles the Pancharatnam phase \cite{pan56} and the other which resembles the phase of Rytov and Vladimirskii \cite{vla41}. Each of these terms depends on the global basis used; but their sum, the geometric phase, is independent of basis. The result of Hannay \cite{han98} follows simply and concisely from one particular choice of global (but not smooth) basis.
The bases used in \cite{ryt38, vla41,  bha89, tav00, cw86, tc86} are all local, not global. We define general local bases and discuss two particular cases of interest, in section 6.

\section{Construction of a globally smooth basis}

We begin with the pair $(\btheta\,, \bvarphi)\,$ of unit vectors along the orthogonal directions of increasing $(\theta\,, \varphi)\,$.  This basis is ill--defined at the north and south poles  but smooth everywhere else.  The following sequence of basis pairs are constructed and their properties listed below, with the final one being the required globally smooth pair.
\begin{eqnarray}
\bfl_{N1} &\;=\;& \cos{\varphi}\,\btheta \;-\; \sin{\varphi}\,\bvarphi\,,
\nonumber\\[1ex]
\bfl_{N2} &\;=\;& \sin{\varphi}\,\btheta \;+\; \cos{\varphi}\,\bvarphi\,.
\label{eqn_ln}
\end{eqnarray}

\noindent
This linearly polarized pair is smooth at the north pole since the transformation compensates for the counterclockwise rotation of the 
$(\theta\,, \varphi)$ directions there. However, this doubles the clockwise  rotation of $(\btheta\,, \bvarphi)$ at the south pole, and hence this basis is singular at $\theta = \pi\,$. The linearly polarized pair
\begin{eqnarray}
\bfl_{S1} &\;=\;& +\cos{\varphi}\,\btheta \;+\; \sin{\varphi}\,\bvarphi\,,
\nonumber\\[1ex]
\bfl_{S2} &\;=\;& -\sin{\varphi}\,\btheta \;+\; \cos{\varphi}\,\bvarphi\,.
\label{eqn_ls}
\end{eqnarray}

\noindent
is, similarly, smooth at the south pole and singular at the north.
For each of the three linear bases $\{(\btheta\,, \bvarphi)\,; (\bfl_{N1}\,, \bfl_{N2})\,; (\bfl_{S1}\,, \bfl_{S2})\}$ there is a corresponding circular
basis. These are defined by
\begin{eqnarray}
\bfc_R &\;=\;& \frac{1}{\sqrt{2}}\left(\btheta \;+\; \im\bvarphi\right)\,,
\nonumber\\[1ex] 
\bfc_L &\;=\;& \frac{1}{\sqrt{2}}\left(\btheta \;-\; \im\bvarphi\right)\,; 
\label{eqn_c}\\[2em]
\bfc_{NR} &\;=\;& \frac{1}{\sqrt{2}}\left(\bfl_{N1} \;+\; \im\bfl_{N2}\right)
\;=\; \exp{[+\im\varphi]}\bfc_R\,,
\nonumber\\[1ex] 
\bfc_{NL} &\;=\;& \frac{1}{\sqrt{2}}\left(\bfl_{N1} \;-\; \im\bfl_{N2}\right)
\;=\; \exp{[-\im\varphi]}\bfc_L\,; 
\label{eqn_cn}\\[2em]
\bfc_{SR} &\;=\;& \frac{1}{\sqrt{2}}\left(\bfl_{S1} \;+\; \im\bfl_{S2}\right)
\;=\; \exp{[-\im\varphi]}\bfc_R\,,
\nonumber\\[1ex] 
\bfc_{SL} &\;=\;& \frac{1}{\sqrt{2}}\left(\bfl_{S1} \;-\; \im\bfl_{S2}\right)
\;=\; \exp{[+\im\varphi]}\bfc_L\,. 
\label{eqn_cs}
\end{eqnarray}

\noindent
Notice that all the circular bases have well defined state of polarisation everywhere but have phase singularities.  The $(\bfc_R\,, \bfc_L)$ basis has  singular phase at both the poles, while the N and S sets are smooth at N and S  respectively, but  are singular at the opposite poles, S and N. 

The globally smooth basis is given in terms of 
$(\bfc_R\,, \bfc_L)$ by 
\begin{eqnarray}
\bfg_1 &\;=\;& A\exp{[+\im\varphi]}\left\{\cos^2{(\theta/2)}\,\bfc_R 
\;-\; \sin^2{(\theta/2)}\,\bfc_L\right\}\,,\nonumber\\[1ex]
\bfg_2 &\;=\;& A\exp{[-\im\varphi]}\left\{\sin^2{(\theta/2)}\,\bfc_R 
\;+\; \cos^2{(\theta/2)}\,\bfc_L\right\}\,,
\label{eqn_gdefc}
\end{eqnarray}

\noindent
with the normalization factor  $A = 
\left[\cos^4{(\theta/2)} + \sin^4{(\theta/2)}\right]^{-1/2}$. The
$(\bfg_1\,, \bfg_2)$ basis is constructed  to  tend to $(\bfc_{NR}\,, \bfc_{NL})$  
at the north pole and $(-\bfc_{SL}\,, \bfc_{SR})$ at the south pole (note the switch in order in the latter case), and is hence smooth everywhere
on the $S^2$ of directions. Expressing  $(\bfg_1\,, \bfg_2)$ in terms of $(\btheta\,, \bvarphi)\,$  
\begin{eqnarray}
\bfg_1 &\;=\;& B\exp{[+\im\varphi]}\left\{\cos{\theta}\,\btheta \;+\; \im\bvarphi\right\}\,,\nonumber\\[1ex]
\bfg_2 &\;=\;& B\exp{[-\im\varphi]}\left\{\btheta \;-\; \im\cos{\theta}\,\bvarphi\right\}\,,
\label{eqn_gdeftp}
\end{eqnarray}

\noindent
with the normalization factor given by $B = 
\left[1 + \cos^2{\theta}\right]^{-1/2}\,$. This form brings out a simple  physical interpretation - it is readily seen that the basis vector $\bfg_1$ is  right circularly polarized at the north pole of the sphere of directions. As we move southwards, it becomes elliptically polarized with axial ratio $\cos{\theta}$ and the major axis along the $\bvarphi$ direction, thus  linearly polarized on and along the equator and  reversing the sense of traversal of the ellipse in the southern hemisphere and becoming left circular, with a unique well--defined phase that is independent of $\varphi$ at the south pole.
 this is exactly the pattern of polarization and phase (though not intensity) which would be radiated by an electric dipole rotating in the positive sense in the $xy$ plane, viewed as a function of direction.  The second basis vector $\bfg_2$  is left circular at the north pole and elliptical in the northern hemisphere with its major axis along the $\btheta$ direction; it reverses sense, retaining the same major axis, as it evolves to become right circular at the south pole --- one can think of it as the electric field  radiated by a magnetic dipole rotating in the negative sense in the $xy$ plane. 

Complementing  the above physical realisation, we offer another simple way of appreciating the existence of globally smooth nonvanishing complex vector fields on a sphere. Both  $\sin{\theta}\,\btheta)$ and $\sin{\theta}\,\bvarphi$ are  smooth real vector fields on the sphere, vanishing at the south and north poles and winding once around each to make up the total index (Euler characteristic) of two. Such fields can of course be constructed for any pair of antipodal poles by rotation. One then manufactures  a nonvanishing complex vector field simply by ensuring that these antipodal nulls  are not the same for the real and imaginary part.
It is straightforward to normalize one such field, and to construct the other member of theth orthonormal  pair by interchanging real and imaginary parts and changing the sign of any one of them (this corresponds to turning the major axis of a general elliptical polarization by $\pi /2$ and reversing the sense, steps which preserve smoothness). Our choice, (\ref{eqn_gdeftp}), has the real part of $\bfg_1$ vanishing on the 
$y$ axis (of direction space) and the imaginary part vanishing on the $x$ axis. The above choice of basis is far from unique. As discussed in the next section, any other global basis is related to $(\bfg_1\,, \bfg_2)$ through a $U(2)$ valued field on the direction sphere. So all smooth choices of bases belong 
to a single homotopy class.  

\section{Consequences of the global basis}

The globally smooth basis provides us with two complex vector fields 
$\bfg_1(\bfk)$ and $\bfg_2(\bfk)$ on the base $S^2$ of directions. 
The complex electric vector of any light wave propagating along the $\bfk$ direction can be written as 
\beq
\bfE \;=\; z_1\,\bfg_1(\bfk) \;+\; z_2\,\bfg_2(\bfk)\,,
\label{edef}
\eeq

\noindent
where $z_1$ and $z_2$ are any two complex numbers. Hence the six dimensional space of light beams with all directions, polarizations, phases and intensities is $\,S^2\times \mathbb{C}^2\,$, which is the (Cartesian) product of a sphere and a two dimensional complex vector space.\footnote{The fact that making the tangent bundle on the sphere complex also  makes it trivial, is well--known to some mathematicians we have consulted.} This means that once the smooth basis field is given, the triplet $(\bfk\,, z_1\,, z_2)\,$, giving direction and two complex amplitudes, specifies  the plane light wave completely.  The intensity of the beam is 
\beq
\bfE^*\cendot\bfE \;=\; \vert z_1\vert^2 \;+\; \vert z_2\vert^2\,,
\label{int}
\eeq

\noindent 
because $\bfg_1^*\cendot \bfg_1 = \bfg_2^*\cendot \bfg_2 = 1$ and 
$\bfg_1^*\cendot \bfg_2 = \bfg_2^*\cendot \bfg_1 = 0\,$. The same beam can be described in a rotated basis which is also globally smooth. We write the new basis as 
\beq
\left(\bfg'_1(\bfk)\,, \,\bfg'_2(\bfk)\right) = \left(\bfg_1(\bfk)\,, \,\bfg_2(\bfk)\right)\msM^{-1}(\bfk)\,, 
\label{rotbasis}
\eeq

\noindent
where $\msM(\bfk)$ is an $U(2)$ valued field on $S^2\,$. The new complex amplitudes are given by corresponding $U(2)$ rotations, $(z'_1\,, z'_2)^T = 
\msM(\bfk) (z_1\,, z_2)^T\,$, which preserves the form of the intensity given 
in (\ref{int}). 

Beams of non--zero intensity have $\bfE^*\cendot\bfE  > 0\,$. The space describing these beams is $\,\mcT = S^2\times \mathbb{C}^2\setminus\{{\bf 0}\}\,$, where $\mathbb{C}^2\setminus\{{\bf 0}\}$ is the set of all $(z_1, z_2)$ except for the origin $(0, 0)\,$. For a given direction $\bfk\,$, two points, 
$(z_1\,, z_2)$ and $(z'_1\,, z'_2)$, have the same polarization state (or ellipse) if $z'_2/z'_1 = z_2/z_1\,$. Hence the natural complex coordinate
on the Poincar\'e sphere is $\zeta = z_2/z_1\,$, which includes the point at infinity when $z_1 = 0$ and the phase of $z_2$ is irrelevant to the polarization state. This relationship is defined by the Hopf map from $\mathbb{C}^2\setminus\{\bf 0\}$ to $S^2$, given by
\beq
f: \mathbb{C}^2\setminus\{{\bf 0}\} \;\to\; S^2\,,\quad\qquad (z_1, z_2) 
\mapsto z_2/z_1\,.
\label{hopf}
\eeq 

\noindent
Polar coordinates, $\left(\theta_{\rm p}, \varphi_{\rm p}\right)$, 
can be defined by $\zeta = \tan(\theta_{\rm p}/2)\exp(\im\varphi_{\rm p})\,$. 
Then the Stokes parameters (in units of the intensity) for fully polarized 
light correspond to the components of the unit vector $\bfs\,$: 
\begin{eqnarray}
s_x &\;=\;& \frac{z_1^*z_2 + z_2^*z_1}{z_1^*z_1 + z_2^*z_2} 
\;=\; \sin\theta_{\rm p}\cos\varphi_{\rm p}\,;\nonumber\\[1ex]
s_y &\;=\;& -\im\,\left(\frac{z_1^*z_2 - z_2^*z_1}{z_1^*z_1 + z_2^*z_2}\right)\;=\; \sin\theta_{\rm p}\sin\varphi_{\rm p}\,;\nonumber\\[1ex]
s_z &\;=\;& \frac{z_1^*z_1 - z_2^*z_2}{z_1^*z_1 + z_2^*z_2} \;=\; 
\cos\theta_{\rm p} \,. 
\label{stokes}
\end{eqnarray}

\noindent
The ordered pair of unit vectors $(\bfk\,, \bfs)$ determines uniquely a light beam with direction $\bfk$ and polarization state labelled by $\bfs\,$. Therefore the four dimensional manifold combining the direction and state of polarization of light is $\mcL = S^2\times S^2$ the product of two spheres. It should be noted that a given point $\bfs$ may correspond to different polarizations states for different $\bfk$. If we want to know the parameters (axis ratio, handedness and orientation) of the polarization ellipse, we require the information contained in the basis field $\left(\bfg_1(\bfk)\,, \bfg_2(\bfk)\right)\,$.

\section{Parallel--transport of $\bfE$}

\subsection{Fibre bundle describing light beams of non--zero intensity}
Here we construct a fibre bundle with total space $\mcT$ and base $\mcL$, with projection given by using the Hopf map (\ref{hopf}). The projection $\pi_{\rm g}$ requires a choice of basis, which could be any of the global bases which are related to each other through the $U(2)$ valued field $\msM(\bfk)$, as discussed earlier; this is indicated by the subscript ${\rm g}\,$. Then there are global coordinates $(\bfk, z_1, z_2)$ on $\mcT$. Under the action of $\pi_{\rm g}\,$, there are global coordinates on $(\bfk, \zeta = z_2/z_1)$ on $\mcL\,$. We define parallel--transport on $\mcT$, using Pancharatnam's principle \cite{pan56} of comparing phases of light beams of different polarizations through interference. Since we are dealing with two beams in infinitesimally close directions, we first have to carry out parallel trannsport on the sphere of directions to bring the second beam to the same direction as the first, consistent with Vladimirskii \cite{vla41}  The result  may be stated simply: \emph{$\bfE$ is said to be parallel--transported when its local change in phase is zero; i.e. the projection of each complex field onto the previous one in the sequence is in phase with it}.  The notion of parallel--transport in $\mcT$ is formalized as 
an $U(1)$ connection on the fibre bundle. In the construction and use of this connection we follow the framework presented by Samuel and Bhandari \cite{sb88}.  In  the case treated by them,  the total space is the set of (nonvanishing) states of a Hilbert space, which is not so here, but the construction goes through without any change.

The fibre bundle $(\mcT\,, \pi_{\rm g}\,, \mcL\,, \mcF\,, A)$ consists of the following elements:
\begin{enumerate}
\item The \emph{total space} is the set of all nonzero intensity light beams $\mcT = S^2\times \mathbb{C}^2\setminus\{{\bf 0}\}\,$, defined by
\beq
\mcT \;=\; \{(\bfk, \bfE)\;\vert\; \bfk\cendot\bfE = 0\,,\quad \bfE^*\cendot\bfE  > 0\,\}\,.
\label{normspace}
\eeq

\item The \emph{base space} $\mcL = S^2\times S^2$ is the set of all
light beams of all directions and polarizations. 

\item The (typical) \emph{fibre} is $\mcF\sim \mathbb{C}\setminus\{0\}\,$, the 
complex plane minus the origin. This set of non--zero complex numbers is 
contractible to the unit circle $S^1\in\mathbb{C}$, so that $\mcF$ is the set 
of all complex numbers to which a unique phase (modulo $2\pi$) can be assigned. 

\item The \emph{projection} $\pi_{\rm g}$ from the total space to the base depends on the global basis chosen; the subscript ${\rm g}$ refers to the  
global basis $\left(\bfg_1(\bfk)\,, \,\bfg_2(\bfk)\right)$ of (\ref{eqn_gdeftp}). Any of the other global bases discussed earlier can also be used. For instance, $\pi_{{\rm g}'}$ can refer to the projection map when the basis
$\left(\bfg'_1(\bfk)\,, \,\bfg'_2(\bfk)\right)$ is used. Henceforth we assume that a  choice of the global basis has been made. Then $\bfE$ is given by the ordered pair of complex numbers $(z_1, z_2)\,$ as in (\ref{edef}), so that $(\bfk, z_1, z_2)$ are global coordinates on $\mcT\,$. The 
Hopf map of (\ref{hopf}) can be used to define the projection we want:
\beq 
\pi_{\rm g}: \mcT\to \mcL\,,\quad\qquad (\bfk, z_1, z_2)\mapsto (\bfk, \zeta = z_2/z_1)\,.
\label{proj}
\eeq

\item The inverse image, 
\beq
\pi^{-1}_{\rm g}(\bfk, \zeta) \;=\; \{(\bfk, z_1, z_2)\in\mcT\;\vert\; z_2/z_1 = \zeta\,\}\,,
\label{invimag}
\eeq

\noindent
is the fibre over the base point $(\bfk, \zeta)\,$, which is diffeomorphic to
the (typical) fibre $\mcF\sim \mathbb{C}\setminus\{0\}\,$. Moving along any fibre corresponds to changing both the intensity and phase of a given polarization state. Changes of phase are of interest to us, so 
we imagine following the changes on the unit circle to which the fibre
is contractible. Therefore, at fixed direction and intensity, the 3--sphere of polarizations and phases is realized as the total space of a non--trivial Hopf bundle with phase as $S^1$ fibre over the Poincare sphere  $S^2$ of polarizations (see Nityananda \cite{nit79} for an elementary treatment of this). 

\item 
Given any curve $(\bfk(\ell)\,, \bfE(\ell)$) in the total space $\mcT\,$, with a real parameter $\ell\,$, the tangent vector to the curve is 
$\left(\rmd\bfk/\rmd\ell\,, \rmd\bfE/\rmd\ell = \bfU(\ell)\right)$. We define 
\beq
A_{\ell} \;=\; \frac{{\rm Im}\{\bfE^*\cendot\bfU\}}{\bfE^*\cendot \bfE} 
\label{con}
\eeq 

\noindent 
as the contraction of a connection one--form on the tangent vector, and 
will refer to it henceforth as the \emph{connection} for brevity. Physically, 
$A_{\ell}$ measures the rate of change of phase along the curve, and the sign is so chosen that a positive value corresponds to phase lagging as $\ell$ increases. We note some salient properties of $A_{\ell}$ below: 
\begin{enumerate}
\item $A_{\ell}$ is a linear functional on the tangent space at any 
point of $\mcT\,$. The definition given above is independent of basis.

\item Of the two components, $\rmd\bfk/\rmd\ell$ and $\bfU(\ell)$, only the latter contributes to $A_{\ell}\,$. This is because  $A_{\ell}$ measures the rate of change of the phase of $\bfE$ with respect to the parameter $\ell\,$, as is evident from (\ref{gamindep});  phase changes are insensitive to the principal--curvature, $\rmd\bfk/\rmd\ell\,$, of the space curve corresponding to the curve $\bfk(\ell)$ in direction space. 

\item Gauge transformations move $\bfE$ along the fibre: $\bfE(\ell) \to 
\rho(\ell)\exp{\left[i\alpha(\ell)\right]}\bfE(\ell)\,$ with $\rho(\ell) > 0\,$.
Since $A_{\ell}$ measures rate of change of phase with respect to $\ell\,$, it is independent of $\rho(\ell)$ which measures change in magnitude. Hence $A_{\ell}$ is a $U(1)$ connection on the fibre bundle, which transforms inhomogenously as:
\beq
A_{\ell} \;\to\; A_{\ell} \;+\; \frac{\rmd\alpha}{\rmd\ell}\,.
\label{gaugetr}
\eeq

\end{enumerate}
\end{enumerate}

\subsection{The $U(1)$ connection}

Interference between two fully polarized beams $\bfE_1$ and $\bfE_2$ traveling in the same direction gives,
\beq
\parallel \bfE_1 \;+\; \bfE_2\parallel^2 \;\;=\;\; \bfE_1^*\cendot \bfE_1 \;+\; \bfE_2^*\cendot \bfE_2 \;+\; 2{\rm Re}\{\bfE_1^*\cendot\bfE_2\}\,.
\label{panch}
\eeq
Pancharatnam \cite{pan56} noted that $\bfE_1$ and $\bfE_2$, referring to 
different non--orthogonal polarization states, can be said to be `in phase'
when, for fixed intensities of the individual beams, the total intensity of the superposition is maximized. Thus the phase difference is taken to be zero  when $\{\bfE_1^*\cendot\bfE_2\}$ is real and positive. More generally, he defined the phase lag of  $\bfE_1$  with respect to $\bfE_2$ 
to be the argument of $\bfE_1^*\cendot\bfE_2\,$. We (and the earlier authors cited)  apply the same concept  to the case where the  electric field vectors lie in infinitesimally close planes rather than in the same plane.  The component of the electric field vector associated with  the new direction perpendicular to the original plane is first order in the angle between the planes, but it does not contribute to the scalar product. 

Let the curve, $(\bfk(\ell)\,, \bfE(\ell)$), in the total space $\mcT$
represent a light beam that is taken around the curve, $\bfk(\ell)\,$, in direction space. \emph{The electric field $\bfE(\ell)$ is defined to be parallel--transported when it changes least.} By this we mean that $\bfE(\ell)$ is in phase with $\bfE(\ell + \rmd\ell)\,$, while $\bfk(\ell)\cendot\bfE(\ell) = \bfk(\ell + \rmd\ell)\cendot\bfE(\ell +\rmd\ell)=0 $. Therefore  parallel transport implies that ${\rm Im}\{\bfE^*\cendot\bfU\} = 0\,$. We derive the $U(1)$ connection and discuss its consequences, using methods given in Samuel and 
Bhandari \cite{sb88}: 

\begin{enumerate}
\item
A curve in $\mcT$ on which $A_{\ell} = 0$ is defined to be a \emph{horizontal curve}. A horizontal  curve in  $\mcT$ is said to be a \emph{horizontal lift} 
of its projection onto the base.

\item
The horizontal lift of a closed curve in $\mcL$ may be open in $\mcT$; i.e the initial and final points lie on the same fibre but have different phases and magnitudes. The phase difference is the geometric phase, formally described as the (an)holonomy of the $U(1)$ connection on the base.

\item
Let $\left(\bfk(\ell)\,, \bfE(\ell)\right)$ be a horizontal curve in $\mcT$ such that, as $\ell$ varies from $0$ to $1$, the light beam returns to the same direction and polarization state. This means
\beq
\bfk(1) \;=\;  \bfk(0)\,,\qquad\quad\bfE(1) \;=\; \lambda\bfE(0)\,,
\label{return}
\eeq

\noindent
where $\lambda$ is a non zero complex number. The initial and final states
are on the fibre over the same base point, and the final lags the initial by an amount ${\rm Arg}\{\lambda\}$, which is the geometric phase. 

\item
Define a closed curve $C_{h}$ by completing the circuit in $\mcT$ by joining $\left(\bfk(1)\,, \bfE(1)\right)$ to $\left(\bfk(0)\,, \bfE(0)\right)$ by a segment running along the fibre. It can be verified that
\beq
\gamma \;=\; \oint_{C_h} A_{\ell}\,\rmd\ell \;=\; -{\rm Arg}\{\lambda\}
\;+\; 2n\pi\,, 
\label{gamma}
\eeq

\noindent
where $n$ is the winding number of the (vertical) segment running along the fibre. This is because, along the horizontal curve $A_{\ell} = 0\,$, and all the contribution comes from the vertical segment which is equal to $-{\rm Arg}\{\lambda\} + 2n\pi\,$.\footnote{Since the fibre $\mathbb{C}\setminus\{{\bf 0}\}$ is contractible to $S^1$, curves from a given initial to a final point on the same fibre can differ by an integral number of windings around this $S^1$.} Note that $\gamma$ as defined in (\ref{gamma}) is minus the conventional geometric phase, which is the phase of $\bfE(1)\,$ with respect to $\bfE(0)\,$. Henceforth we refer to $\gamma$ as the \emph{geometric phase}.

\item 
$C_h$ projects to a closed curve $C$ in the base $\mcL = S^2\times S^2\,$, which is the locus of points $\left(\bfk(\ell)\,, \zeta(\ell)\right)\,$, where 
\beq
\bfk(1) \;=\;  \bfk(0)\,,\qquad\quad\zeta(1) \;=\; \zeta(0)\,.
\label{return2}
\eeq

\noindent
The integral in (\ref{gamma}) is along a specific closed curve $C_h$ in the total space $\mcT$ which projects to a closed curve $C$ on the base. However, it is gauge invariant, because $A_{\ell}$ transforms according to (\ref{gaugetr}). Therefore $\gamma$ is a property of $C$ (up to an integral multiple of $2\pi$), making the geometric phase factor $\exp{[\im\gamma]}$ 
a property of the closed curve $C$.
\end{enumerate}
 
\section{The geometric phase}

Let $C$ be a closed curve on $\mcL = S^2\times S^2$ with parameter $\ell$ 
that varies from $0$ to $1\,$. Let $\tilde{C}$ be  a closed path in $\mcT$, 
obtained by lifting $C$. Gauge invariance implies that the geometric phase is given by, 
\beq
\gamma \;=\; \oint_{\tilde{C}} A_{\ell}\,\rmd\ell \;=\; 
\oint_{\tilde{C}}\frac{{\rm Im}\{\bfE^*\cendot\bfU\}}{\bfE^*\cendot \bfE} \,\rmd\ell\,.
\label{gamindep}
\eeq

\noindent
Thus $A_{\ell}$ measures rate of change of phase, along the curve 
$\tilde{C}\,$, with respect to the parameter $\ell\,$. This expression also makes it clear that the geometric phase is independent of the basis used.
Just like in (\ref{gamma}), the value of the integral in (\ref{gamindep}) can differ by integral multiples of $2\pi\,$, when it is evaluated along paths in $\mcT$ that are both lifts of $C$ but not homotopic to each other. Once a basis is chosen, $\mcL$ acquires coordinates: these could be the ordered pair $(\bfk, \zeta)$, or $(\bfk, \bfs)$ where $\bfs$ is the Stokes vector. 

\subsection{Globally smooth basis}

We write the electric field in the global basis 
$\left(\bfg_1(\bfk)\,, \bfg_2(\bfk)\right)$ as
\beq
\bfE(\ell) \;=\; z_1(\ell)\,\bfg_1(\bfk(\ell)) \;+\; z_2(\ell)\,\bfg_2(\bfk(\ell))\,,
\label{edefell}
\eeq
and define the state vector,
\beq
\vert\psi\rangle \;=\; \left(\begin{array}{c}
		z_1(\ell)\\
		z_2(\ell)
		\end{array}\right)\,.
\label{ket}
\eeq

\noindent
which is an element of $\mcN\sim\mathbb{C}^2\setminus\{{\bf 0}\}$, 
the nonzero states of a two dimensional Hilbert space $\mcH\,$.
When $z_1$ and $z_2$ return to their original values along the closed 
path $\tilde{C}$, their arguments $\xi_1$ and $\xi_2$ can change only by integral multiples of $2\pi\,$:
\beq
\xi_1(1) \;\;=\;\; \xi_1(0) \;+\; 2n_1\pi\,,\qquad
\xi_2(1) \;\;=\;\; \xi_2(0) \;+\; 2n_2\pi\,,
\label{intchange}
\eeq

\noindent
where $n_1$ and $n_2$ are integers. Since $z_2/z_1 = \tan(\theta_{\rm p}/2)\exp(\im\varphi_{\rm p})\,$, we have $\varphi_{\rm p} = (\xi_2-\xi_1)$,
and the change in $\varphi_{\rm p}$ along the path $C$ is:
\beq
\varphi_{\rm p}(1) \;\;=\;\; \varphi_{\rm p}(0) \;+\; 2(n_2 - n_1)\pi\,,
\label{phipchange} 
\eeq

\noindent
so $(n_2 - n_1)$ measures the winding number of the azimuth $\varphi_{\rm p}$ on the Poincar\'e sphere, and is hence a property of the curve $C$ on the base $\mcL\,$. For a given path on the base --- i.e keeping $(n_2-n_1)$ fixed --- $n_1$ counts winding on the fibre, and different values correspond to homotopically distinct lifts $\tilde{C}$ of the path on the base. We find it useful to keep the explicit  dependence on both these integers (which is commonly left implicit) in the discussion which follows.

Substituting (\ref{edefell}) in  (\ref{con}), we obtain,
\beq
A_{\ell} \;=\; \frac{{\rm Im}\left\{\langle\psi\vert(\rmd/\rmd\ell)\vert\psi\rangle\right\}}{\langle\psi\vert\psi\rangle}
 \;+\; 
\frac{\langle\psi\vert\,\hat{H}\,\vert\psi\rangle}{\langle\psi\vert\psi\rangle}\,,
\label{aexpr}
\eeq

\noindent
where $\hat{H}$ is a Hermitian operator --- henceforth referred to as a `Hamiltonian' --- which defined on $\mcH$ (and hence $\mcN\,$), whose matrix elements are given by,
\beq
H_{ij} \;=\; \frac{1}{2\im}\left(\bfg_i^*\cendot \frac{\rmd\bfg_j}{\rmd\ell} \;-\; \frac{\rmd\bfg_i^*}{\rmd\ell}\cendot\bfg_j\right)\,.
\label{hamdef}
\eeq 

\noindent
Of the two terms on the right side of (\ref{aexpr}), the first can be written
(upto a total derivative) in terms of coordinates on the Poincar\'e sphere. 
Direct computation gives 
\beq 
\frac{{\rm Im}\left\{\langle\psi\vert(\rmd/\rmd\ell)\vert\psi\rangle\right\}}{\langle\psi\vert\psi\rangle}\,\rmd\ell \;\;=\;\; \case{1}{2}\left(1 - \cos\theta_{\rm p}\right)\rmd\varphi_{\rm p} \;+\; \rmd\xi_1\,,
\label{panchcon}
\eeq

\noindent
The second term of (\ref{aexpr}) can be thought of as the expectation value of $\hat{H}$ in the state $\vert\psi\rangle\,$. It depends on both the rate of change of the basis along the curve, as well as the polarization state.

The total geometric phase can now be written as
\beq 
\gamma \;=\; \gamma_{\rm P} \;+\; \gamma_{\rm RV}\,,
\label{geomph}
\eeq

\noindent
the sum of two quantities. The first is a Pancharatnam--like (hence P) phase: 
  
\begin{eqnarray}
\gamma_{\rm P} &\;=\;& \oint_{\tilde{C}}\frac{{\rm Im}\left\{\langle\psi\vert(\rmd/\rmd\ell)\vert\psi\rangle\right\}}{\langle\psi\vert\psi\rangle}\,\rmd\ell\nonumber\\[1em] 
&\;=\;& \case{1}{2}\oint_{C}\left(1 - \cos\theta_{\rm p}\right)\rmd\varphi_{\rm p} \;+\; \oint_{\tilde{C}}\rmd\xi_1\nonumber\\[1em]
&\;=\;& \case{1}{2}\Omega_{\rm p} \;+\; 2n_1\pi\,,
\label{gp}
\end{eqnarray}

\noindent
where $\Omega_{\rm p}$ is the solid angle on the Poincar\'e sphere swept
by the shortest meridional arc that connects the north pole to the Stokes vector $\bfs\,$, and the contribution of $2n_1\pi$ comes from the total change in $\xi_1$ over the closed path $\tilde{C}$.\footnote{Note that $\Omega_{\rm p}$ changes discontinuously by $\pm 4\pi$ when the contour crosses the south pole, contributing an additional amount $\pm\case{1}{2}4\pi = \pm 2\pi$ to $\gamma_{\rm P}$, which is harmless because it does not change the geometric phase factor $\exp{[\im\gamma]}$.} Even though the value of $\gamma_{\rm P}$ is basis--dependent, its form is invariant, and equal to 
$\case{1}{2}\Omega_{\rm p} + 2n_1\pi\,$. We refer to the  second term, 
\beq
\gamma_{\rm RV} \;=\; \oint_{\tilde{C}}\frac{\langle\psi\vert\,\hat{H}\,\vert\psi\rangle}{\langle\psi\vert\psi\rangle}\,\rmd\ell\,,
\label{gr}
\eeq

\noindent
as a Rytov--Vladimirskii (hence RV) phase, since they  derived it for a situation when there are changes of direction but  polarization remains circular.  The phase  $\gamma_{\rm RV}$ is more general than \cite{ryt38, vla41} in that it also keeps track of  changes of  polarization. The
elements of the Hamiltonian $\hat{H}$ can be computed by substituting 
(\ref{eqn_gdeftp}) in (\ref{hamdef}):
\begin{eqnarray}
H_{11}\,\rmd\ell &\;=\;& -H_{22}\,\rmd\ell \;=\; \frac{\sin^2{\theta}}{1 \,+\, \cos^2{\theta}}\,\rmd\varphi \,,
\nonumber\\[1em]
H_{12}\,\rmd\ell &\;=\;& H^*_{21}\,\rmd\ell \;=\;
-\,\frac{\exp{[-2\im\varphi]}\sin{\theta}}{2\left(1 \,+\, \cos^2{\theta}\right)}
\left[\sin{2\theta}\,\rmd\varphi \,+\, 
2\im\,\rmd\theta\right]\,.
\label{eqn_hamglobal}
\end{eqnarray}

\noindent
Note that the use of the global basis gives a Hamiltonian which is smooth
everywhere: terms singular at the poles are multiplied by factors which vanish there. Using this is (\ref{gr}), it is straightforward to calculate the RV phase as a line integral over the closed curve $C$ on the base:
\begin{eqnarray}
\gamma_{\rm RV} &\;=\;& \oint_{C}\frac{\sin^2{\theta}\left[\cos{\theta_{\rm p}} \,-\, 
\cos{\theta}\sin{\theta_{\rm p}}
\cos{\left(\varphi_{\rm p} - 2\varphi\right)}\right]}
{1 \,+\, \cos^2{\theta}}\,\rmd\varphi
\nonumber\\[1em]
&&\qquad +\; \oint_{C}\frac{\sin{\theta}\sin{\theta_{\rm p}}
\sin{\left(\varphi_{\rm p} - 2\varphi\right)}}
{1 \,+\, \cos^2{\theta}}\,\rmd\theta\,.
\label{grglobal}
\end{eqnarray}

\noindent
The total geometric phase, $\gamma = \gamma_{\rm P} + \gamma_{\rm RV}$, is 
obtained by adding the expressions in (\ref{gp}) and (\ref{grglobal}). 
However, the separation should not be taken literally, since both $\gamma_{\rm P}$ and $\gamma_{\rm RV}$ depend on the basis chosen. It is only their sum $\gamma$ which is independent of basis.  

We can also take account of geometric phases due to lumped elements introduced in the path of the light ray, causing abrupt changes of polarization. Let the  state vector change by a finite amount, from $\vert\psi\rangle$ to $\vert\psi'\rangle$, when $\ell$ changes by an infinitesimal amount $\rmd\ell\,$. Since the change in direction is smooth, the change in $\bfk$ is also infinitesimal; the integrand of (\ref{gr}) being finite, the contribution to $\gamma_{\rm RV}$ during this jump in polarization is of order $\rmd\ell$, and can be neglected. However, $\gamma_{\rm P}$ gets a finite contribution, which is given by the \emph{geodesic rule} of \cite{sb88}: the phase change is given by adding a line segment $\Delta\tilde{C}$ to the integral of (\ref{gp}), such that $\Delta\tilde{C}$ is a geodesic curve in $\mcN$, connecting $\vert\psi\rangle$ and $\vert\psi'\rangle\,$. This projects to a geodesic, $\Delta C\,$, which is a great circle on the Poincar\'e sphere, connecting the initial and final Stokes vectors, $\bfs$ and $\bfs'\,$. Hence, phase changes due to multiple lumped elements encountered by a light ray can be accounted for, by adding geodesic arcs on the Poincar\'e sphere (along which evolution is continuous) to the curve $C$ in $\mcL\,$. 

\subsection{Global but singular basis}

Whereas the expression in (\ref{grglobal}) for $\gamma_{\rm RV}$ is smooth 
in the global basis, it is difficult to interpret physically, because the global basis is variable on the direction sphere. However, we can now  utilize the freedom  to  use a  non--global basis set for computation of any quantity whose existence is guaranteed by a globally smooth basis. Using the circular basis $\left(\bfc_{NR}\,, \bfc_{NL}\right)$ of (\ref{eqn_cn}) means dealing with  the familiar azimuthal phase singularity at the south pole of our coordinate system. 
Let the state vector in this basis be $\vert CN\rangle = (z'_1\,, z'_2)^T\,$. It is straightforward to calculate the Hamiltonian $\hat{H}_{CN}$ as,
\beq
\hat{H}_{CN} \;=\; (1 - \cos\theta)\,\frac{\rmd\varphi}{\rmd\ell}
\left( \begin{array}{cc}
		\;1 \;&\; \,0 \\[1ex]
		\,0 \;&\; -1
		\end{array}\right)\,,
\label{hcn}
\eeq 

\noindent
Using this in (\ref{aexpr}), the connection one--form is,
\begin{eqnarray}
A_{\ell}\,\rmd\ell &\;=\;&
\frac{{\rm Im}\left\{\langle CN\vert(\rmd/\rmd\ell)\vert CN\rangle\right\}}{\langle CN\vert CN\rangle} \;+\; 
\frac{\langle CN\vert\,\hat{H}_{CN}\,\vert CN\rangle}{\langle CN\vert CN\rangle}\nonumber\\[1em]
&\;=\;& \case{1}{2}\left(1 - \cos\theta_{\rm p}\right)\rmd\varphi_{\rm p} 
\;+\; \rmd\xi_1 \;+\; \cos\theta_{\rm p} (1-\cos\theta)\,\rmd\varphi\,,
\label{acn}
\end{eqnarray}

\noindent
which is, as expected, singular at the south pole ($\theta = \pi$) of
the direction sphere. Integrating, the geometric phase
is given by the compact expression,
\beq
\gamma \;\;=\;\; \oint_{\tilde{C}} A_{\ell}\,\rmd\ell
\;\;=\;\; \case{1}{2}\Omega_{\rm p} \;+\; 2n_1\pi \;+\; 
\oint_{C}\cos\theta_{\rm p} (1-\cos\theta)\,\rmd\varphi\,,
\label{gammacn}
\eeq

\noindent
as the sum of a P phase and an RV phase. We have seen that the former
has an invariant form. The RV phase has the following structure: the basis being circular, $\cos\theta_{\rm p} = V$, the Stokes $V$ parameter which 
is a measure of the degree of circular polarization; and $(1-\cos\theta)\,\rmd\varphi$ is equal to the change in the solid angle traced by the shortest meridional arc connecting the north pole to the contour on the direction sphere (the $CN$ basis being singular at the south pole, we assume that the contour does not traverse the south pole). Note that the physical meaning of the separation into P and RV phases, is specific to the chosen (i.e. $CN$) basis. We will see more examples of such basis--dependent separation in the next section.

The work of Hannay \cite{han98}, using the Majorana representation for spin one, is formulated in terms of vectors in $\mathbb{R}^3$ (ordinary three space),  and gives a concise one--term formula for the geometric phase (equation (16)  in \cite{han98}). This formula is manifestly intrinsic, i.e it is expressed in a form  which makes it clear that it is independent of the basis chosen in polarization and direction spaces.  We now place  this  result in our framework. Working in the $CN$ circular basis as earlier,
we write the complex electric field as $\bfE = z_1\bfc_{NR} + z_2\bfc_{NL}$. We now make the gauge choice  $\xi_1 = -\varphi_{\rm p}/2$; since $\varphi_{\rm p} = \left(\xi_2 - \xi_1\right)$, this implies that $\xi_2 = \varphi_{\rm p}/2$, so that 

\beq\fl\qquad
z_1 \;=\; \vert\bfE\vert\cos(\theta_{\rm p}/2)\exp[-\im\varphi_{\rm p}/2]\,,\qquad
z_2 \;=\; \vert\bfE\vert\sin(\theta_{\rm p}/2)\exp[+\im\varphi_{\rm p}/2]\,. 
\label{z12han}
\eeq

\noindent
It can be verified that the zero of phase is at the tip of the major axis of the ellipse: i.e. at time $t=0$ the real electric field vector points along
the major axis which itself makes an angle of $\varphi_{\rm p}/2$ with 
$\bfl_{N1}\,$. This choice of phase, apart from being undefined at both the poles of the Poincar\'e sphere, runs into a discontinuity of $\pi$ in phase as we go around the sphere once in the azimuthal direction, since the ellipse turns by $180^{\circ}$ and the zero of phase is transferred to the opposite end of the major axis. Therefore it is necessary to introduce a branch cut which could be chosen along a line of constant longitude joining the two poles. However, neither of these affect the application of this gauge in the small. The formula for the connection can be obtained by setting 
$\xi_1 = -\varphi_{\rm p}/2$ in (\ref{acn}):
\beq
A_{\ell}\,\rmd\ell \;\;=\;\; -\case{1}{2}\cos\theta_{\rm p}\,\rmd\varphi_{\rm p} \;+\; \cos\theta_{\rm p}(1-\cos\theta)\rmd\varphi
\;\;=\;\; -\cos\theta_{\rm p}\,\delta\beta\,,
\label{acn1}
\eeq

\noindent
where $\delta\beta = \left[\case{1}{2}\rmd\varphi_{\rm p} - (1-\cos\theta)\rmd\varphi\right]$ has the physical interpretation of an intrinsic or covariant change in the azimuth of the major axis, with the second term allowing for the rotation of the $\left(\bfl_{N1}, \bfl_{N2}\right)$ frame under parallel displacement along the $\bvarphi$ direction, because there is no rotation when we move along $\btheta\,$. Equation (\ref{acn1}) is equivalent to equation (16) of \cite{han98}, so we see that the results of that paper follow from a particular basis choice in our framework, and are not dependent on using the Majorana representation. The cautionary remarks following equation (16) in \cite{han98} follow from the existence of the branch--cut in this basis (which the curve $C$ may traverse).

\section{The geometric phase in local bases}

The bases used in \cite{ryt38, vla41,  bha89, tav00, cw86, tc86} are all of a local character; i.e. they are specified once the curve in parameter space is given, but not in advance. The bases also differ from each other, so that the expressions for $\gamma_{\rm P}$ and $\gamma_{\rm RV}$, obtained by different authors, take different forms. It is straightforward to write the geometric phase when a \emph{local basis}, $\left(\bfe_1(\ell)\,, \,\bfe_2(\ell)\right)\,$, is used. The local basis is defined on the curve, $\bfk(\ell)\,$ by, $\left(\bfe_1(\ell)\,, \,\bfe_2(\ell)\right) = \left(\bfg_1(\bfk(\ell))\,, \,\bfg_2(\bfk(\ell))\right)\msR^{-1}(\ell)\,$, where $\msR(\ell)$ is a $U(2)$ valued field which is defined on the curve in direction space.\footnote{The local basis can be extended, if necessary, to a tube of trajectories surrounding the curve $\bfk(\ell)\,$.} This basis is orthonormal: $\bfe_1^*\cendot \bfe_1 = \bfe_2^*\cendot \bfe_2 = 1$ and $\bfe_1^*\cendot \bfe_2 = \bfe_2^*\cendot \bfe_1 = 0\,$. The new complex amplitudes are $(w_1\,, w_2)^T = \msR(\bfk) (z_1\,, z_2)^T\,$. Then the complex electric field can be written as, 
\beq
\bfE(\ell) \;=\; w_1(\ell)\,\bfe_1(\ell) \;+\; w_2(\ell)\,\bfe_2(\ell)\,.
\label{edefcirc}
\eeq

\noindent
Using this in (\ref{aexpr}), the connection one--form is:
\beq
A_{\ell} \;=\; \frac{{\rm Im}\left\{\langle\phi\vert(\rmd/\rmd\ell)\vert\phi\rangle\right\}}{\langle\phi\vert\phi\rangle}
 \;+\; 
\frac{\langle\phi\vert\,\hat{K}\,\vert\phi\rangle}{\langle\phi\vert\phi\rangle}\,,
\label{aexprloc}
\eeq

\noindent
where $(w_1(\ell)\,, w_2(\ell))^T = \vert\phi(\ell)\rangle = \msR(\ell)\vert\psi(\ell)\rangle$ and $\hat{K}$ is the new Hamiltonian $\hat{K}$, whose components are,
\beq
K_{ij} \;=\; \frac{1}{2\im}\left(\bfe_i^*\cendot \frac{\rmd\bfe_j}{\rmd\ell} \;-\; \frac{\rmd\bfe_i^*}{\rmd\ell}\cendot\bfe_j\right)\,.
\label{kam}
\eeq

\subsection{Local circular basis} 

A local circular basis satisfies the usual orthonormality conditions, $\bfc_1^*\cendot \bfc_1 = \bfc_2^*\cendot \bfc_2 = 1$ and $\bfc_1^*\cendot \bfc_2 = \bfc_2^*\cendot \bfc_1 = 0\,$. The Hamiltonian
\beq
\hat{K}_{\rm circ} \;=\; \frac{1}{2\im}\left[\bfc_1^*\cendot \frac{\rmd\bfc_1}{\rmd\ell} \;-\; \bfc_2^*\cendot\frac{\rmd\bfc_2}{\rmd\ell}\right]
\left( \begin{array}{cc}
		\;1 \;&\; \,0 \\[1ex]
		\,0 \;&\; -1
		\end{array}\right)\,,
\label{kamcirc}
\eeq 

\noindent
is diagonal because, under a parallel displacement the circular basis 
vectors do not mix; each can only acquire a phase.\footnote{We have already encountered an example of this in (\ref{hcn}), for the circular basis $\left(\bfc_{R}\,, \bfc_{L}\right)\,$.} It is straightforward to calculate the connection one--form:
\beq
A_{\ell}\,\rmd\ell \;=\; \case{1}{2}\left(1 - \cos\theta_{\rm p}\right)\rmd\varphi_{\rm p} \;+\; \rmd\xi_1 \;+\;
\frac{V}{2\im}\left(\bfc_1^*\cendot \rmd\bfc_1 \;-\; \bfc_2^*\cendot\rmd\bfc_2\right)\,,
\label{acirc}
\eeq

\noindent
where $V = \cos\theta_{\rm p}$ is the Stokes $V$ parameter, measuring the degree of circular polarization along the curve. We can express the second term of (\ref{acirc}) in terms of properties intrinsic to the space curve. Begin with the linearly polarized basis used by Rytov \cite{ryt38}: $(\bfn(l), \bfb(l))\,$, where $\bfn(l)$ is the normal and $\bfb(l)$ is the binormal to the space curve described by the light ray; the parameter $l$  measures length along the space curve. Define the \emph{Rytov circular basis} as,
\beq
\bfc^{\rm r}_R (l) \;=\; \frac{1}{\sqrt{2}}\left[\bfn(l) \,+\, \im\bfb(l)\right]\,,\qquad
\bfc^{\rm r}_L (l) \;=\; \frac{1}{\sqrt{2}}\left[\bfn(l) \,-\, \im\bfb(l)\right]\,,
\label{rytbas}
\eeq

\noindent 
and use the Frenet--Serret formulae to get, $(2\im)^{-1}\left(\bfc^{{\rm r}*}_1\cendot \rmd\bfc^{\rm r}_1 \;-\; \bfc^{{\rm r}*}_2\cendot\rmd\bfc^{\rm r}_2\right) = -\tau(l)\rmd l$, where $\tau(l)$ is the torsion of the space curve. Then the geometric phase is, 
\beq
\gamma \;=\; \oint_{\tilde{C}} A_{l}\,\rmd l \;=\; 
\case{1}{2}\Omega^{\rm r}_{\rm p} \;+\; 2n_1\pi
\;\;-\;\; \oint V(l)\tau(l)\,\rmd l\,,
\label{gamryt}
\eeq

\noindent
as before the sum of two phases. The second is the line integral of the product of Stokes V and torsion $\tau\,$, a property intrinsic to the space curve described by the light ray. For circularly polarized light, when $V= \pm 1\,$,  we encounter the integral of the torsion over a closed curve which equals $\left(2\pi - \Omega\right)$, consistent with the known result for  this case. The expression given above in (\ref{gamryt}) resembles (\ref{gammacn}) for the geometric phase in the $CN$ basis (which is global but singular). However, the former has the advantage of using quantities intrinsic to the curve. Circular bases are used in \cite{cw86, tc86, bb87}.  

We address the concern that the normal and binormal are undefined for a straight section, or at a point of inflection,  of our space curve. If the curve happens to be entirely straight, any orthogonal pair of real vectors in the transverse plane would suffice. The case of interest is therefore when part of the space curve has a well--defined curvature and normal. Working with  the corresponding  path on the sphere of directions, we see that each point represents a unit tangent $\bfk$ to the space curve. The unit tangent to the path on the sphere is just the normal $\bfn$ to the original space curve. A straight section is then just a single point, but the curve must continue in order to return to its starting point and, for a smooth continuation, there is no difficulty in specifying $\bfn$, and hence the binormal $\bfb = \bfk\cross\bfn\,$,  in a continuous manner.

\subsection{Parallel--transported basis}

Any basis, chosen at an initial point $l = 0\,$, can be evolved by 
parallel displacement. A useful property of an initial circular basis is 
that it always remains circular, acquiring only phase as it is evolved
along the curve by parallel displacement. We take the initial basis at $l = 0$ to agree with the  Rytov circular basis, but undo the phase change which it undergoes by parallel displacement. Therefore, from (\ref{gamryt}), we see that the pair of circular states defined by,
\beq\fl\qquad
\bfp_1(l) \;=\; \exp{\left[+\im\,\mbox{$\int_0^l\tau(l')\,\rmd l'$}\right]}\bfc^{\rm r}_1(l)\,,\qquad
\bfp_2(l) \;=\; \exp{\left[-\im\,\mbox{$\int_0^l\tau(l')\,\rmd l'$}\right]}\bfc^{\rm r}_2(l)\,,
\label{parbas}
\eeq

\noindent
is a parallel--transported circular basis; \cite{bha89, tav00} use bases
like these. Unlike the bases used so far, the initial (at $l=0$) and final (at $l=1$) basis vectors are not the same. The vectors acquire a phase of minus (for right circular) and plus (for left circular) the solid angle $\Omega$ in direction space: 

\beq
\bfp_1(1) \;=\; \exp{\left[-\im\Omega\right]}\,\bfp_1(0)\,,\qquad
\bfp_2(1) \;=\; \exp{\left[+\im\Omega\right]}\,\bfp_2(0)\,.
\label{parbas2}
\eeq

\noindent
When the complex electric field,
\beq
\bfE(l) \;=\; w_1(l)\,\bfp_1(l) \;+\; w_2(l)\,\bfp_2(l)\,,
\label{edefpt}
\eeq

\noindent
returns to the same fibre, $\,\bfE(1) = \lambda\bfE(0)\,$, the initial and 
final amplitudes are given by
\beq
w_1(1) \;=\; \lambda\exp{\left[+\im\Omega\right]}\,w_1(0)\,,\qquad
w_2(1) \;=\; \lambda\exp{\left[-\im\Omega\right]}\,w_2(0)\,.
\label{w12pt}
\eeq

\noindent
As discussed in item~(v) of section~4.2, we have the freedom to evaluate the integral on a convenient closed curve in the total space, so we can set 
$\lambda =1$ in (\ref{w12pt}). If $\xi_1$ and $\xi_2$ be the arguments of $w_1$ and $w_2$, we have 
\beq 
\xi_1(1) \;=\; \xi_1(0) \;+\; \Omega \;+\; 2n_1\pi\,,\qquad\;\;
\xi_2(1) \;=\; \xi_2(0) \;-\; \Omega \;+\; 2n_2\pi\,,
\label{phasespt}
\eeq

\noindent
where $n_1$ and $n_2$ are integers. Then, 
\beq
\frac{w_2(1)}{w_1(1)} \;=\; \exp{\left[-\im\,2\Omega\right]}\, 
\frac{w_2(0)}{w_1(0)}\,,
\label{wratio}
\eeq

\noindent
implies that the curve $C_{\rm op}\,$ on the Poincar\'e sphere (associated with the complex  coefficients $(w_1, w_2)$ in the local basis) is, in general, not closed; its end points have equal latitudes but different longitudes:
\beq
\theta_{\rm p}(1) \;=\; \theta_{\rm p}(0)\,,\qquad 
\varphi_{\rm p}(1) \;=\; \varphi_{\rm p}(0) \;-\; 2\Omega
\;+\; 2(n_2 - n_1)\pi\,.
\label{open}
\eeq

\noindent
We emphasize that this auxiliary Poincar\'e sphere, associated with the local basis, should not be confused with the sphere of polarization in a global basis, which occurs in the definition of the $S^2\times S^2$ base space; on this latter sphere the path is of course closed. 

\noindent
\emph{When (\ref{parbas}) is substituted (\ref{kam}), the Hamiltonian $\hat{K}$ vanishes; this is the characteristic property of the parallel--transported basis}. Hence the connection one--form is given by just the Pancharatnam term,
\beq 
A_l\,\rmd l \;=\; \frac{{\rm Im}\left\{\langle\phi\vert(\rmd/\rmd l)\vert\phi\rangle\right\}}{\langle\phi\vert\phi\rangle}\rmd l \;\;=\;\; 
\case{1}{2}\left(1 - \cos\theta_{\rm p}\right)\rmd\varphi_{\rm p} \;+\; \rmd\xi_1\,.
\label{panchin}
\eeq

\noindent
Integrating, and using (\ref{phasespt}), the geometric phase is: 
\begin{eqnarray}
\gamma &\;=\;& \int_{\tilde{C}} A_l\,\rmd l \;=\;
\case{1}{2}\int_{C_{\rm op}}(1 - \cos\theta_{\rm p})\,\rmd\varphi_{\rm p} 
\;+\; \left[\xi_1(1) \,-\, \xi_1(0)\right]\nonumber\\[1em]
&\;=\;& \case{1}{2}\Omega^{\rm pt}_{\rm p} \;+\; \Omega \;+\;2n_1\pi\,,
\label{gampt}
\end{eqnarray}

\noindent
where $\Omega^{\rm pt}_{\rm p}$ is the solid angle on the Poincar\'e sphere, 
swept by the shortest meridional arc joining the north pole to the Stokes vector, as the latter traverses the open curve $C_{\rm op}$ whose end points are given by (\ref{open}). Equation (\ref{gampt}) is a compact and pleasing form: apart from the usual $2n_1\pi$ term, the geometric phase is given as the sum of one--half of the swept--out  solid angle on the Poincar\'e sphere and the solid angle on the direction sphere. For right circular light, the solid angle on the Poincar\'e sphere $\Omega^{\rm pt}_{\rm p} = 0$, and the geometric phase is simply $(\Omega + 2n_1\pi)$, as it should be. However, the use of this formula calls for a limiting process when applied to the case when the polarization remains left circular throughout, since there are multiple geodesics connecting the north pole to the south pole. The degeneracy is lifted by introducing an infinitesimal ellipticity. From (\ref{open}) we see that the initial and final longitudes on the Poincar\'e sphere are well--defined and differ by $\left[-2\Omega\ + 2(n_2 - n_1)\pi\right]$, and the solid angle $\Omega^{\rm pt}_{\rm p}$ swept out by the meridian is twice this amount. Then (\ref{gampt}) gives the correct phase, equal to 
$(-\Omega\ + 2n_2\pi)$ for left circular polarization. 

\section {Conclusions}
 
The manifold describing the family of plane monochromatic light waves with all directions, polarizations, phases and intensities is seen to be 
$\,S^2\times \mathbb{C}^2\,$, the  (untwisted) Cartesian product of a sphere and a two dimensional complex vector space.  The four dimensional manifold, which describes beams with all directions and polarizations, is $\,S^2\times S^2\,$, the Cartesian product of two spheres. The  fibre bundle, whose total space is the set of all light beams of non--zero intensity $\,S^2\times \mathbb{C}^2\setminus\{{\bf 0}\}\,$, with base space $\,S^2\times S^2\,$, has a $U(1)$ connection which allows a basis--independent definition of the geometric phase. When expressed  in a general global basis, the geometric phase decomposes into the sum of two terms,  one of which resembles the Pancharatnam phase, and the other the phase of Rytov and Vladimirskii. Each term depends on the global basis used, but their sum, the geometric phase, is independent of basis. 

The existence of a globally smooth description allows the use of  
non--global bases for computations. The geometric phase assumes a compact and simple form in terms of the polar coordinates on the Poincar\'e and directions spheres.  The final result  for  the geometric phase in Hannay's work \cite{han98} follows, in our framework, from a particular choice of basis (singular at the poles and along a cut joining them). Two examples are given of oppositely circularly polarized bases, which are not defined globally, but only over the path traversed by the system.  The first  follows Rytov in using the normal and binormal to the space curve, and the second uses parallel displacement, both giving interesting forms for the geometric phase. It is intriguing that one seems to be forced to choose between global smoothness and physical transparency: the smoothest form with which we begin is the most opaque, whereas the parallel--displaced local basis (which does not even return to the same value at the end of a closed path in 
direction/polarization space) gives a compact and pleasing result. 

We conclude with the global conceptual difference between  our framework and the widely used  spin one representation. The latter  admits  an inner product between any two states, not just two nearby ones: the total space is $\mathbb{C}^3$, the space of normalized vectors is $S^5$, and the space of states is $CP^2$. We find that the space of normalized vectors for this problem is $S^2\times S^3\,$, which is not a vector space and does not have a globally defined inner product. An example bringing out  this global distinction is a pair of   oppositely circularly polarized waves of unit intensity travelling in opposite directions: they have identical electric fields and hence would have a vector space inner product with a modulus of unity, even though they are not the same state by any means. The $U(1)$ connection given above in (\ref{con}) --- which determines the geometric phase accrued when a closed path is traversed  in $S^2\times S^2$ --- relies only on the inner product between two nearby electric fields. In mathematical terms, the set of all possible  directions and their associated complex transverse electric fields is a vector bundle with a connection, which is not  a vector space. We have presented a framework that describes the global structure of the space of directions and polarizations, and this allows us to view earlier work on the geometric phase for light in a unified way. 

\ack
RN thanks N~Nitsure and V~Srinivas for explanations of the  mathematics, in particular Hirzebruch surfaces, which do  not appear explicitly here, but were  crucial  in forcing us to  look harder  at the manifold $S^2\times S^2\,$. SS thanks Han Mathews for discussions on topology.


\section*{References}

\begin{thebibliography}{99}

\bibitem{ber26}
Bertolotti~E 1926 \emph{Atti R. Accad. Naz. Lincei Rend. Cl. Sci. Fis. 
Mat. Nat.} {\bf 4} 552

\bibitem{ryt38}
Rytov~S~M 1938 Transition from wave to geometrical optics \emph{Dokl. Akad. 
Nauk. USSR} {\bf 18} 238; reprinted in ref.~5, p 6--10 

\bibitem{vla41}
Vladimirskii~V~V 1941 The rotation of polarization plane for curved light ray
\emph{Dokl. Akad. Nauk. USSR} {\bf 21} 222; reprinted in ref.~5, p 11--16 

\bibitem{pan56} Pancharatnam~S 1956 Generalized theory of interference, and its applications \emph{Proc. Ind. Acad. Sci. A} {\bf 44} 247--262; reprinted in ref.~6, p 51

\bibitem{ber84} Berry~M~V 1984 Quantal Phase Factors Accompanying Adiabatic Changes \emph{Proc. R. Soc. Lond. Ser. A} {\bf 392} 45--57

\bibitem{mv89} Markovski~B and Vinitsky~V~I 1989 \emph{Topological Phases in Quantum Theory} (Singapore, World Scientific)

\bibitem{sw89} Shapere~A and Wilczek~F 1989 \emph{Geometric Phases in Physics}
(Singapore, World Scientific)

\bibitem{bha89} Bhandari~R 1989 Geometric phase in an arbitrary evolution of a light beam \emph{Phys. Lett. A} {\bf 135} 240--244

\bibitem{bha97} Bhandari~R 1997 Polarization of light and topological phases
\emph{Phys. Rep.} {\bf 281} 1--64

\bibitem{han98} Hannay~J~H 1998 The Majorana representation of polarization, and the Berry phase of light \emph{J. Mod. Opt.} {\bf 45} 1001--1008

\bibitem{tav00} Tavrov~A~V, Miyamoto~Y, Kawabata~T and Takeda~M 2000 
Generalized algorithm for the unified analysis and simultaneous evaluation
of geometrical spin--redirection phase and Pancharatnam phase in a complex interferometric system \emph{J. Opt. Soc. Am. A} {\bf 17} 154--161

\bibitem{sim83} Simon~B 1983 Holonomy, the quantum adiabatic theorem, and Berry's phase \emph{Phys. Rev. Lett.} {\bf 51} 2167--2170

\bibitem{sb88} Samuel~J and Bhandari~R 1988 General setting for Berry's phase \emph{Phys. Rev. Lett.} {\bf 60} 2339--2342

\bibitem{cw86} Chiao~R~Y and Wu~Y-S Manifestations of Berry's topological 
phase for the photon \emph{Phys. Rev. Lett.} {\bf 57} 933--936

\bibitem{tc86} Tomita~A and Chiao~R~Y  Observation of Berry's topological 
phase by use of an optical fiber \emph{Phys. Rev. Lett.} {\bf 57} 937--940

\bibitem{bb87} Bialynicki--Birula~I and Bialynicka--Birula~Z 1987
Berry's phase in the relativistic theory of spinning particles 
\emph{Phys. Rev. D} {\bf 35} 2383--2387

\bibitem{nit79} Nityananda~R 1979 Impossibility of a continuous phase convention for polarised light \emph{Pramana} {\bf 3} 257--262

\bibitem{maj32} Majorana~E 1932 
\emph{Nuovo Cim.} {\bf 9} 43


\end{thebibliography}
\end{document}